# Intra-operative Optimal Flow Diverter Selection for Intracranial aneurysm treatment

Parmita Mondal[1,2], Mohammed Mahdi Shiraz Bhurwani[1,2,3], Kyle A Williams[1,2], Ciprian N Ionita[1,2,3]
[1]Department of Biomedical Engineering, University at Buffalo, Buffalo, NY 14260
[2]Canon Stroke and Vascular Research Center, Buffalo, NY 14203
[3]Quantitative Angiographic Systems. Artificial Intelligence, Buffalo, NY 14203

## Abstract

**Purpose:**

During intracranial aneurysm (IA) treatment with Diverters (FDs), the device/parent artery diameters ratio may influence the ability of the device to induce aneurysm healing response. Oversized FDs are safer to deploy but may not induce enough hemodynamic resistance to ensure aneurysm occlusion. Methods based on Computational Fluid Dynamics (CFD) could allow optimal device selection but are time-consuming and inadequate for intra-operative guidance. To address this limitation, we propose to investigate a method for optimal FD selection using Angiographic Parametric Imaging (API) and machine learning (ML).

**Materials and Methods:**
We selected 128 pre-treatment angiographic sequences of that were fully occluded at minimum six months follow-up. For each IA, we extracted five API parameters from the aneurysm dome and normalized them to the feeding artery corresponding parameters. We dichotomized the dataset base on the FD/ proximal artery diameter ratio as undersized, if the ratio<1 or if multiple FDs were used and oversized in the other cases. Using area under the receiver operator characteristic (AUROC), single API parameter and ML analysis was used to determine whether API parameters could be used to determine the need for FD under-sizing (i.e., increase flow resistance).

**Results:** In total we identified 51 and 77 cases for the undersized and oversized cohorts respectively.  Single API parameter analysis yielded an inadequate AUROC
~0.5 while machine learning using all five API parameters yielded and AUROC of 0.72±0.02.

**Conclusion:**

Machine learning based on pre-treatment API could be used to optimize FD selection to improve the aneurysm occlusion rates

## Background

Flow diverters (FDs) are relatively new devices used for the treatment of intracranial aneurysms (IA). As shown in **Figure 1**, reproduced from Shapiro et. al,[1] for a deployed FD, the metal coverage across the aneurysm neck is dependent on the device diameter selection relative to the parent artery. Under sizing of the device will require compressing of the device to reach the nominal artery and achieving more coverage of the IA neck. This approach is not risk free, since the device could percolate in the aneurysm neck due to catheter pushing force, with devastating consequences. CFD simulations could provide a solution by recommending whether under sizing of a device is needed, but they are time consuming and may be unsuitable for intraoperative decision. We propose to investigate whether a method based on a routine angiography acquired during the endovascular procedure could be developed to make a real time recommendation whether under sizing of the device is needed in order to achieve full aneurysm occlusion (i.e., healing).

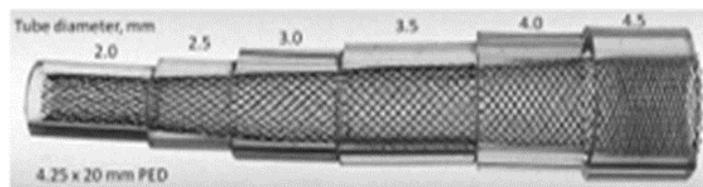

Figure 1: Exemplification of the under sizing versus target artery. A Flow diverter was deployed in a tapered tube. Changes of metal coverage as a function nominal diameter is well exemplified Shapiro et. al.[1]

Digital Subtraction Angiography (DSA) is a technique used in interventional radiology to visualize blood vessels and its surrounding environment. We use the temporal information of the contrast propagation in the blood vessels given by the DSA for better diagnosis.[2,3] The temporal and spatial distribution of the contrast can be useful in performing Angiographic Parametric Imaging (API). This is done by using time- density curves to calculate Time to peak (TTP), Mean transit time (MTT), Time to arrival (TTA), Peak height (PH) and Area Under the Curve (AUC) using an API software. Previous work has demonstrated that the API can be combined with data driven methods to predict aneurysm occlusion immediately after device deployment with a high accuracy.[4] While this approach could have significant clinical application, it still requires a pre- and post-device angiogram, allowing the interventionalist to readjust the treatment only after the first device has been already placed. In this study we propose to extend this approach and determine whether it is possible to make a recommendation for device placement based only on the initial (i.e., pre-device placement) API data.

**Material and Methods**
*Data Collection*

Data collection and analysis was approved by our Institutional Review Board. The **Figure 2** shows chronologically the steps taken to complete this study. DSA scans of n=200 patients were recorded with images before placing the FD (pre-treatment cases) and after placing the FD (post-treatment cases). A follow-up scan of at least 6 months from the placing of treatment were also considered for these images. We used the follow-up to select only the cases that demonstrated full occlusion at six months. The reason for selecting these cases is because the outcome indicates that the device resistance was sufficient to induce aneurysm occlusion. Next, we dichotomized the group based on the FD/artery diameter ratio. If the diameter of the FD is less than the proximal artery, or if multiple FDs were used the case was assigned to the undersized cohort (n=51). The remaining cases (n=77) were assigned to the oversized cohort.

During the actual interventions dozens of (DSA) sequences are acquired for structure assessment. Since the contrast flow obeys the convection equation, the contrast spatial and temporal variability should be directly related to the local vascular hemodynamics. Two-dimensional angiographic parametric imaging API provides a better way to measure hemodynamics by encoding contrast flow parameters from angiograms into a single map that conveys both structure and hemodynamics related imaging biomarkers. Thus, API is a semiquantitative angiographic tool that generates a set of vascular maps, where each pixel is colored according to the intensity of a certain flow parameter. To achieve this, the contrast is monitored in each frame at each pixel to synthesize a time density curve (TDC), which we parametrize to extract the mean transit time (MTT), time to peak (TTP), time to arrival (TTA), peak height (PH) and area under the curve (AUC). This information is used to create a parametrized description of blood flow over the entire vasculature to assess aspects of the flow in a specific lesion, in our case the IA. Each parameter reflects one aspect of hemodynamics information from the angiogram, the combination of which greatly increases the accuracy of the predicted outcome and effects. To reduce hand injection variability, we normalized each parameter to the inlet corresponding values to yield NI-MTT, NI-TTP, NI-TTA, NI-PH and NI-AUC, respectively. Normalized parameters were analyzed in Statistical Package for Social Sciences (SPSS) using standard ROC analysis. T-test was performed for each API parameter using the null hypothesis that: The single API parameter distribution between the two cohorts is the same.

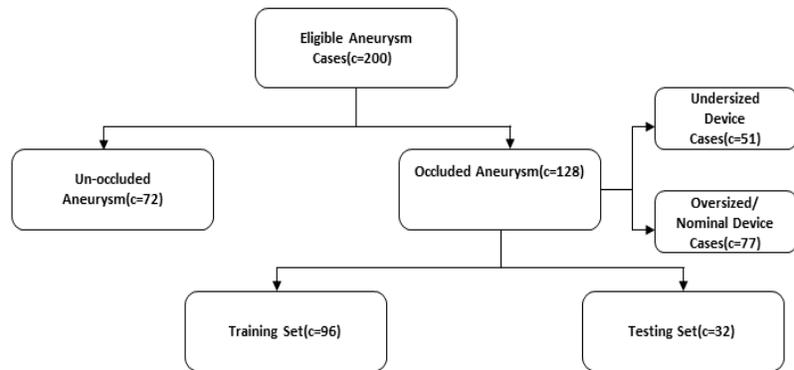

Figure 2: Study inclusion schema for the study (c refers to the number of cases).

A machine learning algorithm based on deep learning neural network (DNN) design was developed in Keras. The model had five layers with 3, 40, 55, 45 and 1 nodes respectively. The input normalized API parameters were:

TTP, MTT, TTA, PH and AUC.[5] The optimizer and loss function are two vital steps in compiling the model. We used 'adam' optimizer and 'mean_squared_error' as our loss function. A fivefold Monte Carlo Cross Validation was used to ensure that poor or excellent performances are not accidental. Accuracy and Area under the Receiver Operating Characteristics (AUROC) were evaluated after each training.

**Results**

API data distribution between the undersized and oversized cohorts is demonstrated using a box and Whisker plot in **Figure 3**. T-test analysis demonstrated that the null hypothesis was retained for every parameter except, NI-TTA and NI-PH with a significance of 0.47 and 0.31 respectively. ROC analysis for single API parameter also demonstrated weak predictive capability for under-sizing the FD. Most parameters were slightly larger than the chance line (**Figure 4**). TTA seemed to have a moderate predictability value with AUROC of 0.604.

The ML algorithm increased tremendously the prediction capabilities. The mean AUROC after fivefold MCCV was 0.72±0.02 with a range of (0.71-0.75), Figure 6.

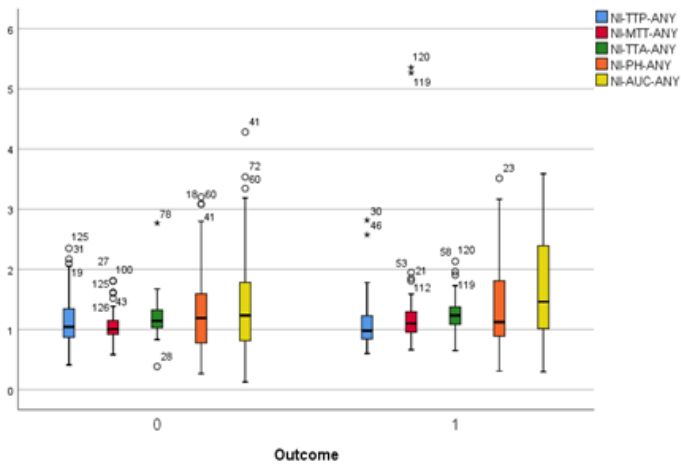

Figure 3: Whisker plots for the API parameters: TTP, MTT, TTA, PH and AUC for (0) oversized and (1) undersized devices (ANY indicate Aneurysm Dome, NI- normalized to inlet)

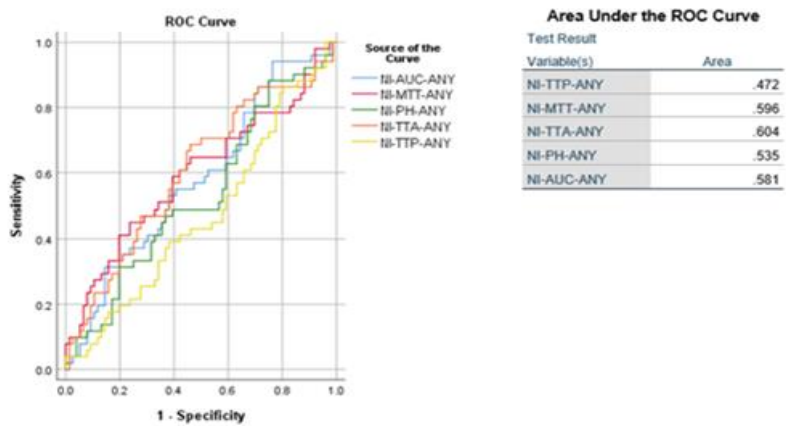

Figure 4: Receiver Operating Characteristics of the API parameters: TTP, MTT, TTA, PH and AUC for the feasibility study

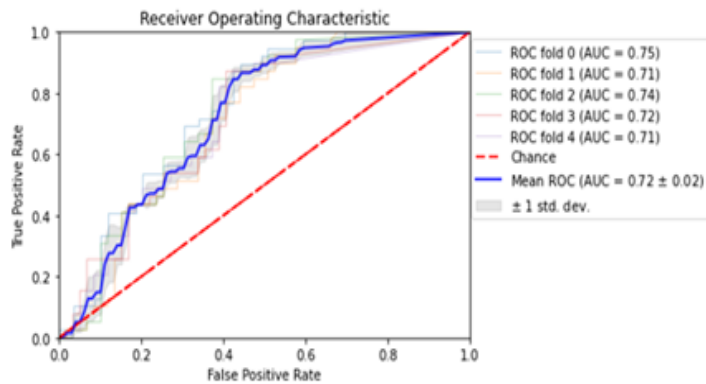

Figure 5: Machine Learning Performance using five-fold Monte Carlo Cross Validation (MCCV)

**Conclusion**

This is a novel attempt to integrate API parameters with DNN to determine correlation between API parameters and hemodynamics and predict the optimal FD selection. The DNNs performed well giving an accuracy of 90% and AUC of 0.72±0.02 when used with normalized data.

**Acknowledgments**

This work was supported by QAS.AI and NSF STTR Award # 2111865